\documentclass[aps, prd, twocolumn, superscriptaddress, nofootinbib]{revtex4-1}
\usepackage{graphicx}
\usepackage{dcolumn}
\usepackage{amssymb}
\usepackage{amsmath,amssymb,amsfonts}

\begin{document}

\newcommand{\Eq}[1]{\mbox{Eq. (\ref{eqn:#1})}}
\newcommand{\Fig}[1]{\mbox{Fig. \ref{fig:#1}}}
\newcommand{\Sec}[1]{\mbox{Sec. \ref{sec:#1}}}

\newcommand{\be}{\begin{equation}}
\newcommand{\ee}{\end{equation}}
\newcommand{\bea}{\begin{eqnarray}}
\newcommand{\eea}{\end{eqnarray}}
\newcommand{\vu}{{\mathbf u}}
\newcommand{\ve}{{\mathbf e}}


\title{Primordial perturbations in a rainbow universe with running Newton constant}

\newcommand{\addressBologna}{Dipartimento di Fisica e Astronomia dell'Universit\`a di Bologna and 
Sez. Bologna INFN, Via Irnerio 46, 40126 Bologna, Italy}
\newcommand{\addressImperial}{Theoretical Physics, Blackett Laboratory, Imperial College, London, SW7 2BZ, United Kingdom}

\author{Francesco Brighenti}
 \affiliation{\addressBologna}
\author{Giulia Gubitosi}
\affiliation{\addressImperial}
\author{Joao Magueijo}
\affiliation{\addressImperial}

\date{\today}

\begin{abstract}
We compute the spectral index of primordial perturbations in a rainbow universe. We allow the Newton constant $G$ to run at (super-)Planckian energies and we consider both vacuum and thermal perturbations.
If the rainbow metric is the one associated to a generalized Horava-Lifshitz dispersion relation, we find that only when $G$ tends  asymptotically to zero can one match the observed value of the spectral index and solve the horizon problem, both for vacuum and thermal perturbations. For vacuum fluctuations the observational constraints imply that the primordial universe expansion can be both accelerating or decelerating, while in the case of thermal perturbations only  decelerating expansion is allowed.
 \end{abstract}


\maketitle

\section{Introduction}

Recent results suggest that the properties of the spectrum of primordial fluctuations might not need inflationary expansion to be explained, but  could instead be a consequence of quantum-gravitational effects, which are relevant in the early universe \cite{Amelino-Camelia:2013tla, Mukohyama:2009gg}. In particular in \cite{Amelino-Camelia:2013tla, Amelino-Camelia:2013wha, Amelino-Camelia:2013gna} it was shown that  a scale invariant power spectrum can be obtained if the perturbations satisfy the Planck-scale-modified dispersion relation emerging in the high-energy regime of Horava-Lifshitz gravity \cite{Horava:2009if}:
\be
E^{2}=p^{2}(1+(\ell p)^{4})\,. \label{eq:2Ddisprel}
\ee
This  dispersion relation implies a running of spacetime dimensionality, so that the spacetime dimension in the deep Planckian regime is 2 \cite{Sotiriou:2011aa, Amelino-Camelia:2013cfa, thermaldimension}. The possibility of generalising this result to any theory with Planck-scale dimensional reduction to 2 was suggested in \cite{gravitybreakdown, Arzano:2015gda}.
These results rely on a number of assumptions, such as that the second order action for perturbations is the one of Einstein gravity and that the perturbations are produced in a quantum vacuum state. This rigidity in the assumptions makes it hard to find a mechanism that would produce the observed small departure from exact scale invariance.

In this paper we shall relax several of the assumptions previously made in the literature. Firstly, we shall 
assume the more general framework of rainbow gravity \cite{Magueijo:2002xx}. The background cosmological evolution will then be described in terms of a metric which ``runs'' with the energy. 
For the dispersion relation:
\begin{equation}
f^2(E)E^2-g^2(E)p^2=m^2,
\end{equation}
(where the continuous functions $f$ and $g$  approach the constant value 1 when the energy is well below the  Planck energy), the associated rainbow line element is
\begin{equation}
ds^2=\frac{dt^2}{f^2(E)}-\frac{1}{g^2(E)}\delta_{ij}dx^idx^j\,.
\end{equation}
Secondly, 
we will consider both perturbations of quantum origin for a vacuum state, and perturbations that are originated in a thermal state \cite{Biswas:2013lna, Ferreira:2007cb, Magueijo:2002pg, Magueijo:2007wf, Koh:2007rx}. In the latter case we will assume that the universe is filled with radiation and that both the background and the fluctuations are thermalized, so that they share the same (modified) thermodynamical properties \cite{Magueijo:2009zp}. Finally, we allow for the Newton constant to also run with energy. This is motivated by  results in Horava-Lifshitz gravity and in Asymptotic Safety \cite{Benedetti:2013pya, DOdorico:2015pil, Lauscher:2001rz, Niedermaier:2006wt}, where the Newton constant tends to zero at super-Planckian energies. We  allow the Newton constant to both increase and decrease with energy. However, it will turn out that in order to solve the horizon problem and to produce perturbations with the required spectral index, the Newton constant must indeed be a decreasing function of energy at super-Planckian scales. This is true for both vacuum and thermal initial conditions for the perturbations.

Regarding our work on thermal fluctuations we note the following motivating factors. 
Radiation obeying a deformed dispersion relation also has deformed thermodynamical properties \cite{thermaldimension, Alexander:2001dr, Santos:2015sva}. In this paper we  focus on a generalisation of the Horava-Lifshitz dispersion relation \eqref{eq:2Ddisprel}:
\be
E^{2}=p^{2}(1+(\ell p)^{2\gamma})\,,\label{eq:MDR}
\ee
and we assume to be in a regime where only the ultraviolet correction term is relevant, $E^{2}\approx p^{2}(\ell p)^{2\gamma}$.
As shown in \cite{thermaldimension}, in this regime the associated Stefan-Boltzmann law and equation of state parameter $w\equiv P/\rho$ are:
\bea
&&\rho\propto   T^{1+\frac{3}{1+\gamma}}    \\
&& w=\frac{1+\gamma}{3} \label{eq:wgamma}\,.
\eea
Using the fact that the equation of state parameter and the Stefan-Boltzmann law for standard radiation depend on spacetime dimensionality as:
\bea
w&=&\frac{1}{d_T-1}\\
\rho&\propto& T^{d_T}
\eea
one can associate the modified disperison relation \eqref{eq:MDR} to the thermodynamical dimension \cite{thermaldimension}
\be
d_{T}=1+\frac{3}{1+\gamma}\,.
\ee

Our paper is structured as follows. In section \ref{sec:background} we start by working  out the evolution of the background, including modified thermodynamical relations. In section \ref{sec:horizon} we derive the equation for the evolution of primordial scalar perturbations and we  derive the constraints on the modified dispersion relation and on the running of the Newton constant which ensure an expanding universe and a solution to the horizon problem. Section \ref{sec:vacuum} is devoted to the computation of the spectral index for perturbations generated in a quantum vacuum, while section \ref{sec:thermal} shows the analogous results for perturbations with thermal initial conditions. We present some conclusions in section \ref{sec:conclusions}. 

\section{Background evolution of a rainbow FLRW universe with deformed thermodynamics} \label{sec:background}

The rainbow functions associated to the dispersion relation \eqref{eq:MDR} are:
\be
f^{2}=1\;\quad g^{2}=1+(\ell p)^{2\gamma}\,.\label{eq:HLrainbowFCT}
\ee
They enter in the rainbow line element for a FLRW spacetime in the following way \cite{Magueijo:2002xx, Santos:2015sva}:
\begin{equation}
ds^2=\frac{dt^2}{f^2(E)}-\frac{a^2(t)}{g^2(E)}\delta_{ij}dx^idx^j.
\end{equation}

We assume that the universe contains a perfect fluid, whose stress-energy tensor is $\mathcal{T}^\mu_\nu=(\rho+P)u^\mu u_\nu-P\delta^\mu_\nu$, where $\rho$ is the energy density, $P$ the pressure and $u^\mu$ the fluid four velocity \footnote{As mentioned in the introduction we also allow for a possible energy dependence of the Newton constant G.}.
Then the Friedmann equations read \cite{Magueijo:2002xx}:
\begin{equation}\label{rainbowBackground}
\begin{split}
H^2&=\frac{8\pi G(E)}{3f^2}\rho\\
H^2-\frac{\ddot{a}}{a}&=\frac{4\pi G(E)}{f^2}(\rho+P),
\end{split}
\end{equation}
where $H=\frac{da/dt}{a}$ .
From these  the continuity equation follows
\begin{equation}\label{continuityequation}
\dot{\rho}=-3H(\rho+P).
\end{equation}
The  solution of the continuity equation can be stated in terms of the equation of state parameter as usual, and if the universe is filled with radiation this translates into a dependence on the parameter $\gamma$ appearing in the dispersion relation \eqref{eq:MDR}:
\be
\rho=\bar{\rho} a^{-3(1+w)}=\bar{\rho} a^{-(4+\gamma)}.
\ee
Of course in the case of standard thermodynamics in four spacetime dimensions $d_T=4$ and we recover the usual scaling $\rho=\bar\rho a^{-4 }$ in a radiation-filled universe.

Using the Stefan-Boltzmann law one finds that the deformed thermodynamics also affects the evolution of the temperature with the scale factor: 
\be
T\propto a^{-3w} = a^{-(1+\gamma)} \,.
\ee

\section{Evolution of scalar perturbations in a rainbow universe and solution to the horizon problem}
\label{sec:horizon}

The perturbed rainbow FLRW metric in the longitudinal gauge\footnote{By this we mean that in the limit where the energy dependence of the metric disappears, $f=g=1$, one is left with the metric in longitudinal gauge.} reads:
\begin{equation}\label{rainbowPert}
ds^2=\frac{dt^2}{f^2(E)}(1+2\phi(t,x))-\frac{a^2(t)}{g^2(E)}(1-2\psi(t,x))\delta_{ij}dx^idx^j.
\end{equation}

In order to work out the evolution equation for the perturbations one can introduce an energy-dependent time variable,
\be
d\tilde{t}=\frac{dt}{f(E)}\,,
\ee 
so that the time-dependent functions appearing in the metric read
\bea
\tilde{a}^2(E,\tilde{t})=\frac{a^2(\tilde{t})}{g^2(E)}, \, \, \, \tilde{\phi}(\tilde{t},x)=\phi(t,x),\, \, \,  \tilde{\psi}(\tilde{t},x)=\psi(t,x)\,.\nonumber\\
\eea
The perturbed line element takes the standard form in terms of the new functions:
\begin{equation}
ds^2=d\tilde{t}^2(1+2\tilde{\phi}(\tilde{t},x))-\tilde{a}^2(E,\tilde{t})(1-2\tilde{\psi}(\tilde{t},x))\delta_{ij}dx^idx^j \,.
\end{equation}
Using these new variables one can just follow a standard procedure (see e.g. \cite{Magueijo:2009zp}) to get the familiar equation for the perturbations:
\begin{equation}\label{pert4}
\tilde{v}''-\left(\nabla^2+\frac{\tilde{z}''}{\tilde{z}}\right)\tilde{v}=0\,,
\end{equation}
where the prime means derivative with respect to the energy-rescaled conformal time, $\frac{d}{d\tilde \eta}\equiv \tilde a(E,\tilde t)\frac{d}{d\tilde t}$, and the scalar perturbation $\tilde v$ is defined as usual as $\tilde v=\tilde \zeta\tilde z$, where  the  curvature perturbation $\tilde{\zeta}$ is 
\be
\tilde{\zeta}=\tilde{\phi}\frac{5+3w}{3(1+w)}+\frac{\tilde{\phi}'}{\tilde{\mathcal{H}}}\frac{2}{3(1+w)}
\ee 
and $\tilde z=\sqrt{\frac{3(1+w)}{2}}\tilde a$. Note that we have set $\tilde c_{s}^{2}\equiv \frac{\delta \tilde P}{\delta \tilde \rho}=1$. 

Going back to the energy-independent time variable one finds that the curvature perturbation is left unchanged,
\be
\tilde{\zeta}=\phi\frac{5+3w}{3(1+w)}+\frac{ad\phi/dt}{da/dt}\frac{2}{3(1+w)}=\zeta\,,
\ee
while 
\be
\tilde{z}=\sqrt{\frac{3(1+w)}{2}}\tilde{a}=\sqrt{\frac{3(1+w)}{2}}\frac{a}{g}=z/g.
\ee
Therefore, $v=\tilde{v} g$ satisfies the following evolution equation in Fourier space
\begin{equation}\label{perturbation}
v''-\left(\frac{g^2}{f^2} k^2+\frac{a''}{a}\right)v=0\,.
\end{equation}
From now on, the prime stands for the derivative with respect to the energy-independent conformal time, $\frac{d}{d\eta}\equiv a\frac{d}{dt}$. This equation is very similar to the standard one, with the factor $(f/g)^{2}$  which plays the role of an energy-dependent speed of sound.

Note that a possible energy dependence of the Newton constant does not appear explicitly in the evolution equations of the perturbations, however we will see in the following that it affects the scale of the horizon and  the conditions under which the horizon problem is solved within rainbow cosmology models.

A cosmological model that solves the  horizon problem is such that modes start inside the horizon, where the first term in parentheses in the evolution equation \eqref{perturbation} dominates, and subsequently exit the horizon, where the second term dominates \cite{Magueijo:2009zp, Noller:2009tj}.
We investigate the conditions under which the horizon problem is solved specialising to  the dispersion relation \eqref{eq:MDR}, with associated rainbow functions \eqref{eq:HLrainbowFCT} and assuming to be in a regime where only the ultraviolet correction terms are relevant. It is important to bear in mind that the energy appearing in the rainbow functions is the physical one,  related to the comoving $k$ via $E=  \left(\frac{\ell k}{a(\eta)}\right)^{2\gamma}$.

The behaviour of the two terms in parenthesis in eq. \eqref{perturbation} is governed by the evolution of the scale factor $a(\eta)$. This is found by integrating the first Friedmann equation  \eqref{rainbowBackground}, leading to
\begin{equation}
\eta^2=\frac{a^{1+3w}}{(1+3w)^2}\frac{1}{\frac{2}{3} \pi \bar{\rho}G} = \frac{a^{2+\gamma}}{(2+\gamma)^2}\frac{1}{\frac{2}{3} \pi \bar{\rho}G} \,.
\end{equation}
Here, $\bar \rho$ is the initial energy density and the realation between the equation of state parameter $w$ and the deformation parameter $\gamma$ is given by the modified thermodynamical relation \eqref{eq:wgamma}.
If the Newton constant is energy-independent, the scale factor evolves as:
\be
a(\eta)=(C \eta^{2})^{\frac{1}{2+\gamma}}\,,
\ee
where $C=G\frac{2}{3} \pi \bar{\rho}(2+\gamma)^2$ and $\eta$ increases from $0$ in order to have cosmological expansion with time.  Then the two terms in parentheses in \eqref{perturbation} take the form
\begin{equation}\label{harm oscill}
k^2 \left(\frac{\ell k}{a(\eta)}\right)^{2\gamma}=k^2(\ell k)^{2\gamma}C^{-\frac{2\gamma}{2+\gamma}}\eta^{-\frac{4\gamma}{2+\gamma}}
\end{equation}
and
\begin{equation}\label{jeans}
\frac{a''}{a}=\eta^{-2}\frac{2}{2+\gamma}\left(\frac{2}{2+\gamma}-1\right).
\end{equation}
The horizon is then found at 
\begin{equation}\label{hor const G}
\eta_H=\left(k^2(\ell k)^{2\gamma}C^{-\frac{2\gamma}{2+\gamma}}\frac{(2+\gamma)^2}{2\gamma}\right)^{\frac{2+\gamma}{2(\gamma-2)}}\,,
\end{equation}
and in order to solve the horizon problem one needs
\be
\gamma >2\,.
\ee

If the Newton constant has a power-law dependence on energy in the ultraviolet regime,
\be\label{UVG}
G(E)=\ell^2(\ell E)^{\alpha}\sim \ell^2\left(\frac{\ell k}{a}\right)^{(1+\gamma)\alpha}\,,
\ee
then the evolution of the scale factor with time is
\begin{equation}
a(\eta)=(\bar C\eta^{2}(\ell k)^{(1+\gamma)\alpha})^{1/\nu} ,
\end{equation}
where $\nu=2+\gamma+(1+\gamma)\alpha$ and $\bar C=\frac{2}{3} \pi \ell^2\bar{\rho}(2+\gamma)^2$. Note that depending on $\nu$ the conformal time can either be positive or negative. In fact, in order to have cosmological expansion with time  if $\nu>0$ then $\eta$ must be positive and increasing from $0$, while if  $\nu<0$ then $\eta$ must be negative and approaching $0$ from $-\infty$.

The  terms in parenthesis in the perturbation equation \eqref{perturbation} are now:
\begin{equation}
k^2 \left(\frac{\ell k}{a(\eta)}\right)^{2\gamma}=\bar C^{-\frac{2\gamma}{\nu}}\eta^{-\frac{4\gamma}{\nu}}k^2(\ell k)^{\frac{2\gamma(2+\gamma)}{\nu}},
\end{equation}
and
\begin{equation}\label{a''/a toy}
\frac{a''}{a}=\frac{2}{\nu}\left(\frac{2}{\nu}-1\right)\eta^{-2}\,.
\end{equation}
The horizon is then found at
\begin{equation}\label{kHor toy}
\eta_{H}=\left(\frac{\nu \,\bar C^{-\frac{2\gamma}{\nu}}}{2\left(\frac{2}{\nu}-1\right)}k^2(\ell k)^{\frac{2\gamma(2+\gamma)}{\nu}}\right)^{\frac{\nu}{4\gamma-2\nu}}
\end{equation}
and the horizon problem is solved for
$
\frac{4\gamma}{\nu}>2
$
 if $\eta$ is positive and for
$
\frac{4\gamma}{\nu}<2
$ 
otherwise. Then the overall conditions on $\alpha$ that ensure cosmological expansion and solution of the horizon problem are 
\begin{equation}\label{condition n pos conf t}
-\frac{2+\gamma}{1+\gamma}< \alpha<\frac{\gamma-2}{1+\gamma}
\end{equation}
for positive $\eta$ and 
\begin{equation}
\alpha<-\frac{2+\gamma}{1+\gamma}\,,\quad \alpha>\frac{\gamma-2}{1+\gamma}
\end{equation}
for negative $\eta$. The latter possibility is obviously excluded. The first option correctly reduces to $\gamma>2$ when $\alpha=0$, while in general it constrains $\alpha$ to be in the range $-2<\alpha<1$.

\section{Vacuum perturbations} \label{sec:vacuum}
We can study the power spectrum of vacuum fluctuations directly in the general case where the UV energy dependence of $G$ is encoded in \eqref{UVG}. The limit $\alpha=0$ gives the results for energy-independent G.

\begin{figure}[h]
\centering
\scalebox{0.65}{\includegraphics{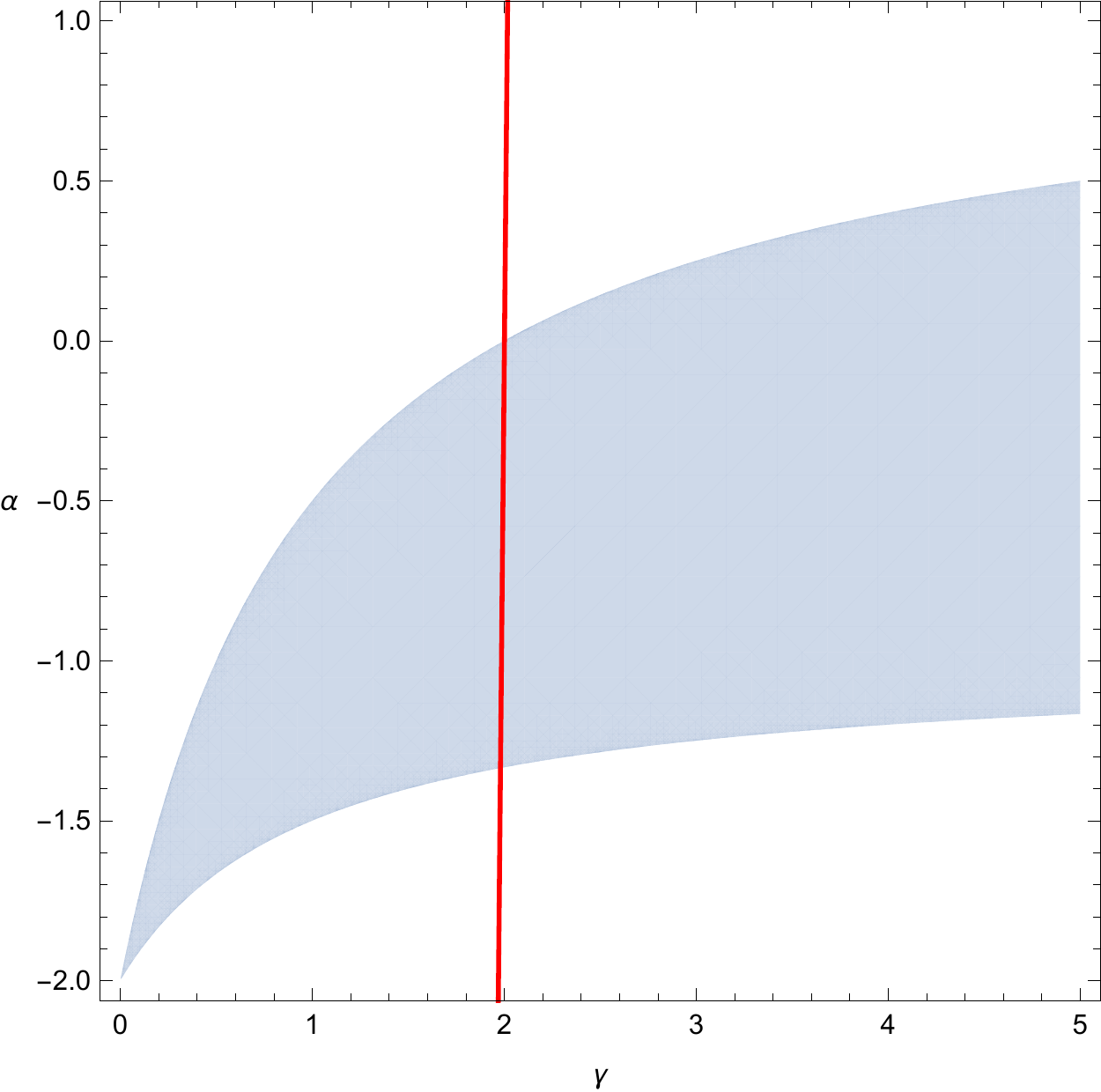}}
\caption{\label{fig:vacuum} We have plotted in red the constraint $n_{s}=0.968\pm0.006$, assuming vacuum fluctuations (the error bar is too small to be seen on the plot). We have plotted in blue the region satisfying the constraint ensuring solution of the horizon problem, eq. \eqref{condition n pos conf t}.}
\end{figure}

The dynamics of modes inside the horizon is governed by the first term in parentheses in \eqref{perturbation}. Up to a phase, the vacuum fluctuations inside the horizon take the form \cite{Amelino-Camelia:2013tla, Amelino-Camelia:2013wha}:
\be
v_{V}\sim  \frac{a^{\gamma/2}}{\sqrt{\ell^\gamma k^{1+\gamma}}}\,.
\ee
The solution of \eqref{perturbation} for modes outside the horizon can be cast in the ansatz:
\be
v_V\sim F(k)a\,,
\ee
where the function $F$ is found by asking that the two solutions match at the horizon:
\begin{equation}\label{hor cross}
F(k)=\frac{a^{\gamma/2-1}(\eta_H)}{\sqrt{\ell^\gamma k^{1+\gamma}}}.
\end{equation}
The dimensionless power spectrum of curvature perturbations $\zeta$ is given by $k^3\mathcal{P}_\zeta\sim k^{3}\left( \frac{v}{z}\right)^{2}\equiv A^{2} k^{n_{s}-1}$. Its spectral index $n_{s}$ is  found from \eqref{hor cross} and \eqref{kHor toy}:
\begin{equation}\label{vacuum runG spectral index}
n_{s}^V-1= \frac{(\gamma+4)(2-\gamma)}{2-\gamma+\alpha(1+\gamma)}\,.
\end{equation}
Clearly $\gamma=2$ gives a scale invariant power spectrum for any value of $\alpha$ allowed by the constraint  \eqref{condition n pos conf t}, which for $\gamma=2$ reads $-\frac{4}{3}<\alpha<0$. The fact that scale invariance is achieved independently of how the Newton constant scales with energy is due to the time perturbations being already scale-invariant and proportional to the scale factor $a$ inside the horizon. So the gluing procedure is trivial, bypassing whatever modified evolution of the background was introduced.  
Also a near-scale invariant power spectrum is allowed. In particular one can ask that $n_{s}^{V}=0.968\pm 0.006$, which is the  present observational constraint from  Planck  \cite{Ade:2015xua}, obtaining the allowed range of values shown in Fig. \ref{fig:vacuum}. Note that now the energy dependence of the Newton constant is relevant. In particular, the values of $\alpha$ that are selected by observational constraints are all negative, suggesting a  vanishing Newton constant in the deep UV regime.
On the other hand, from eq. \eqref{a''/a toy} one can see that observational constraints allow for both an accelerated or decelerated expansion. This is a crucial difference with respect to the constraints coming from thermal fluctuations, as shown in the following section.

In the limiting case $\alpha=0$ (energy-independent Newton constant) the glueing condition at the horizon gives a spectral index which is far from scale invariance, $n_{s}^V-1=4+\gamma$. However, when $\gamma=2$ both the terms governing the evolution of perturbations,  (\ref{harm oscill}) and (\ref{jeans}), scale like $\eta^{-2}$. Therefore a mode is  either inside or outside the horizon, unable to cross it. Whether a mode is inside or outside the horizon is set by the scale
\be
k_H=\left(G\frac{8\pi}{3}\frac{\bar{\rho}}{\ell^4}\right)^{1/6}=H_0\left(\frac{1}{(\ell H_0)^4}\frac{\bar{\rho}}{\rho_{cr}}\right)^{1/6}\,,
\ee
where $H_0$ is the current value of the Hubble constant and $\rho_{cr}$ is the critical energy density. If the modes are well inside the horizon, $k\gg k_H$, the perturbations behave like $v_V\sim \frac{a}{\sqrt{\ell^2 k^3}}$, and so they are  scale-invariant, but never exit the horizon.

\section{Thermal perturbations}\label{sec:thermal}

Without an inflationary phase, there is no real reason to exclude the contribution to the perturbations power spectrum coming from thermalised perturbations, since this is not suppressed by a period of super-cooling~\cite{Magueijo:2007wf, Biswas:2013lna}.
We compute the thermal contribution to the power spectrum applying the method outlined in \cite{Ferreira:2007cb}, but taking into account that in our model both background and perturbations are thermalised.
This in particular means that the same thermodynamical constraints \eqref{eq:wgamma} hold for background and perturbations.
The expectation value of a quantum operator is
\be
\langle O\rangle=\frac{\sum_n \rho_{nn}\langle n|\hat{O}|n\rangle}{\sum_n \rho_{nn}\langle n|n\rangle}\,,
\ee
where $|n\rangle$ is the $n$-particle state. We assume that the density matrix follows the Boltzmann distribution $\rho_{nn}=e^{-\beta E_n}$, where $\beta=1/k_BT$ and $E_n=p_n\sqrt{1+(\ell p_n)^{2\gamma}}$ is the energy of a mode with occupation number $n$.

Then the correlation function of the quantised perturbation variable $\hat v$ is \cite{Magueijo:2007wf}
\begin{equation}
\langle\hat{v}(\vec{x})\hat{v}(\vec{x}+\vec{r})\rangle=\int \frac{d^3k}{(2\pi)^{3/2}}|v_k(\eta)|^2(2n(k,\eta)+1)e^{i\vec{k}\cdot\vec{r}}\,,
\end{equation}
where the number density is given by the Bose-Einstein distribution:
\be
n(k,\eta)=\frac{1}{e^{\beta E(k,\eta)}-1}\,.
\ee 
The power spectrum of thermal perturbations imprinted at the horizon is therefore
\begin{equation}\label{thermalpowerspectrum}
\mathcal{P}_{Therm}(k)=\mathcal{P}_{Vac}(k)(2n(k,\eta_H)+1).
\end{equation}
Since we are in the Rayleigh-Jeans limit for the regime of fluctuations being studied, we can set:
\begin{equation}
n(k,\eta_H)\approx (\beta E)^{-1} = \frac{k_B T_c\ell }{ (\ell k)^{\gamma+1}},
\end{equation} 
where the conformal  temperature $T_c\equiv T a^{\gamma+1}$ is constant in time. As in \cite{Ferreira:2007cb, Agarwal:2014ona}, the relation between the physical and conformal temperature is found by asking that the number density is independent of time. If $c$ is $k$ independent, this is just $T_c=Ta/c$. Here we should strip off the k dependence in $c$ from the definition of $T_c$, so that it does not become $k$ dependent. 

Including the thermal contribution, the spectral index of perturbations becomes
\begin{equation}\label{nTvsnS}
n_s^{T}=n_s^V-1-\gamma.
\end{equation}
Note that this result differs form the one in \cite{Magueijo:2008yk}, because a mistake has been made there. 
In the Rayleigh Jeans limit, $n\sim T/E$, not just $T/k$. The fact that $c$ has an extra dependence in 
$k$ is responsible for the last term in (\ref{nTvsnS}). 
This result is also independent of how the Newton constant runs with energy. 

Using the value of the vacuum spectral index found in the previous section, eq. \eqref{vacuum runG spectral index}, the thermal spectral index can be written as 
\begin{equation}\label{thermal spectral index}
n_s^{T}=\frac{4(2-\gamma)-\alpha \gamma\,(1+\gamma)}{2-\gamma+\alpha(1+\gamma)}\,.
\end{equation}

For energy-independent Newton constant, $\alpha=0$, the thermal spectral index is 
\be
n_{s}^{T}=4\,,
\ee
regardless of the value of $\gamma$. This result matches the one found in \cite{Magueijo:2007wf, Biswas:2013lna} and of course it is ruled out by observational constraints.

For $\alpha\neq0$, asking that the perturbations are scale invariant leads to a constraint linking $\alpha$ and $\gamma$. Asking in addition that the horizon problem is solved, eq. \eqref{condition n pos conf t},  introduces a bound from below on the allowed values of $\gamma$, $\gamma>2$. Then the values of $\alpha$ that are compatible with scale invariance and which allow to solve the horizon problem fall in the range $-1/4<\alpha < 0$. 

\begin{figure}[h]
\centering
\scalebox{0.65}{\includegraphics{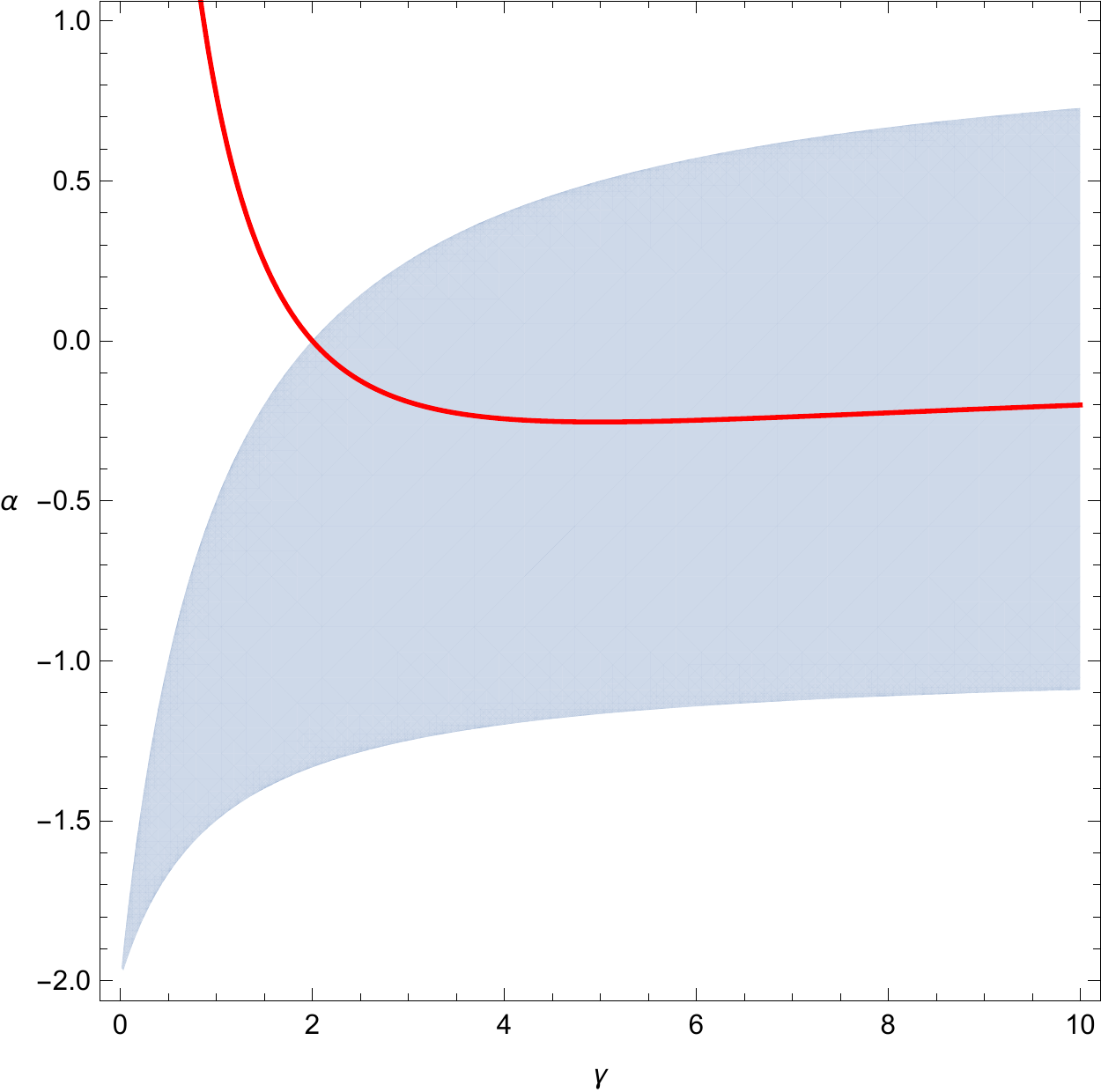}}
\caption{\label{fig:thermal} We have plotted in red the constraint $n_{s}=0.968\pm0.006$, assuming thermal fluctuations (the error bar is too small to be seen on the plot). We have plotted in blue the region satisfying the constraint ensuring a solution of the horizon problem, eq. \eqref{condition n pos conf t}.}
\end{figure}

It is also possible to match the spectral index to the Planck observed value $n_{s}=0.968 \pm 0.006$ \cite{Ade:2015xua}, giving the constraints shown in Fig. \ref{fig:thermal}.
According to eq.(\ref{a''/a toy}), these observational constraint on $\alpha$ and $\gamma$ only allow for a decelerating expansion of the universe.

\section{Conclusions}\label{sec:conclusions}

We have investigated the possibility that a rainbow universe with running Newton constant can accommodate primordial perturbations whose spectral index matches current constraints, without relying on inflation to solve the horizon problem.
Starting form a universe filled with radiation subject to deformed dispersion relations (of the Horava-Lifshitz type), we considered both vacuum and thermal initial conditions for the perturbations and assumed a power-law dependence of the Newton constant on energy. Crucially, we assumed that the background satisfies the thermodynamical relations peculiar to radiation subject to deformed dispersion relations. 

For both kinds of initial conditions for the perturbations (vacuum and thermal) the running of the Newton constant is essential in achieving a viable picture. In particular, the Newton constant is constrained to be decreasing with energy in the ultraviolet regime. This is consistent with intuition from quantum gravity theories, such as Horava-Lifshitz gravity and Asymptotic safety. It also resonates with the conjecture put forward in \cite{gravitybreakdown}. In our scenarios, vacuum and thermal initial conditions can be distinguished because, while for the former the observational constraints are compatible with either an accelerating or decelerating expansion of the universe, for the latter only a decelerated expansion is allowed.

One may question the wisdom of enforcing thermodynamical constraints on the background as well as on the fluctuations. A counter-example is a scalar field, for which the background does not need to be thermalized even when the fluctuations are \cite{Ferreira:2007cb}. Nonetheless it is curious that when, for the sake of minimality, one imposes thermal conditions on both background and perturbations of a scalar field, one recovers the universal result $n_s^T=4$ previously derived for a thermodynamical fluid~\cite{Magueijo:2007wf}. Just as with~\cite{Magueijo:2007wf} one needs to relax standard assumptions to evade this result. Here the running of Newton's constant was the crucial ingredient.

\end{document}